\documentclass[12pt]{article}
\usepackage{pic02}
\usepackage{hyperref}
\usepackage{url}
\usepackage{graphicx}

\begin{document}

\title{\bf THE VALENCE QUARK DISTRIBUTION OF THE PION$^1$}
\author{J. P. Lansberg, F. Bissey, J. R.  Cudell, J. Cugnon, M. Jaminon  and P. Stassart \\
{\em Universit\'e de  Li\`ege, D\'epartement de  Physique B5a,}\\
{\em Sart Tilman, B-4000 LIEGE 1, Belgium} \\
}
\maketitle

\baselineskip=14.5pt
\begin{abstract}\setcounter{footnote}{1}
\footnotetext{Presented by J. P. Lansberg at the
22nd Physics in Collision Conference (PIC 2002), Stanford, California, June 20-22, 2002.}
The pion structure function is investigated in a simple model, where the pion and
its constituent quark fields are coupled through the simplest pseudoscalar
coupling. The imaginary part of the forward $\gamma^{\star} \pi \rightarrow
\gamma^{\star}\pi$ scattering amplitude is evaluated and related to the
structure functions. It is shown that the introduction of non-perturbative
effects, linked to the size of the pion, allows
a connection with the quark distribution. 
It is predicted that higher-twist terms become negligible for $Q^2$ larger than $\sim$2~GeV$^2$,
that quarks in the pion have a momentum fraction smaller than in the proton
case, and that the momentum sum rule is violated for the pion.
\end{abstract}

\baselineskip=17pt

\section{The Model}

We consider~\cite{Bissey:2002yr} an isospin
triplet pion field $\vec{\pi}=(\pi^+,\pi^0,\pi^-)$ interacting with quark
fields $\psi$ through the  Lagrangian density
\begin{equation}
 {\mathcal L}_{int}= i g (\overline{\psi}\mbox{  $\vec{\tau}$} \gamma_5 \psi) .
\mbox{  $\vec{\pi}$}, \label{Lint}
\end{equation}
where $\vec{\tau}$ is the isospin vector operator. 
The imaginary part of the amplitude for $\gamma^{\star} \pi \rightarrow
\gamma^{\star}\pi$ is  considered at lowest order ($\alpha g^2$) and 
obtained using Cutkosky rules. The 
coupling constant $g$ (see Fig.~\ref{figure1}.a) is chosen to fulfill the number of
particles sum rule ($\int F_1(x) dx = \frac{5}{18}$). The effects of the pion 
size are mimicked by the introduction of a 
cut-off ($\Lambda$) on an observable quantity, {\it i.e.} $t$ for the process 
$\gamma^{\star} \pi \rightarrow q \bar q$, --thus in a gauge invariant way--.

\section{Results}

\begin{figure}[htbp]
  \centerline{\hbox{ \hspace{0.2cm}
    \includegraphics[width=5.5cm]{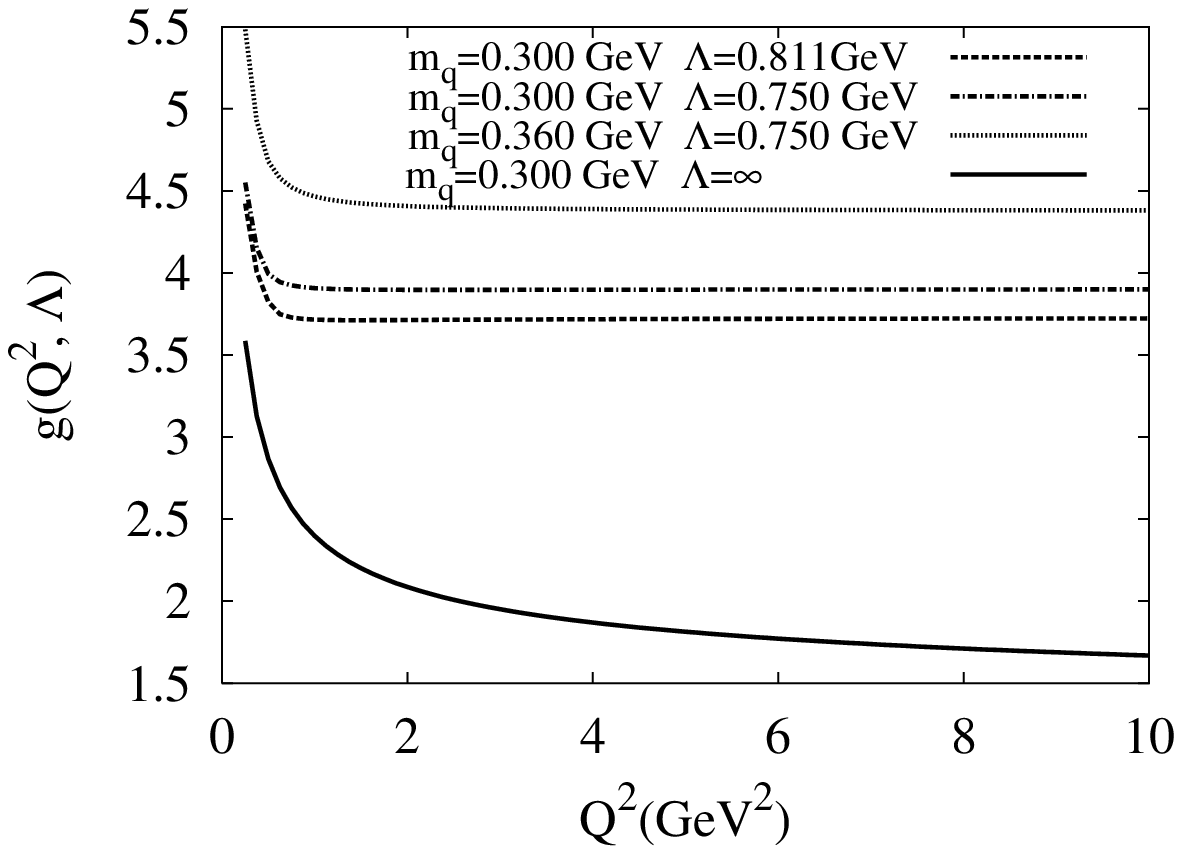}
    \hspace{0.3cm}
    \includegraphics[width=5.5cm]{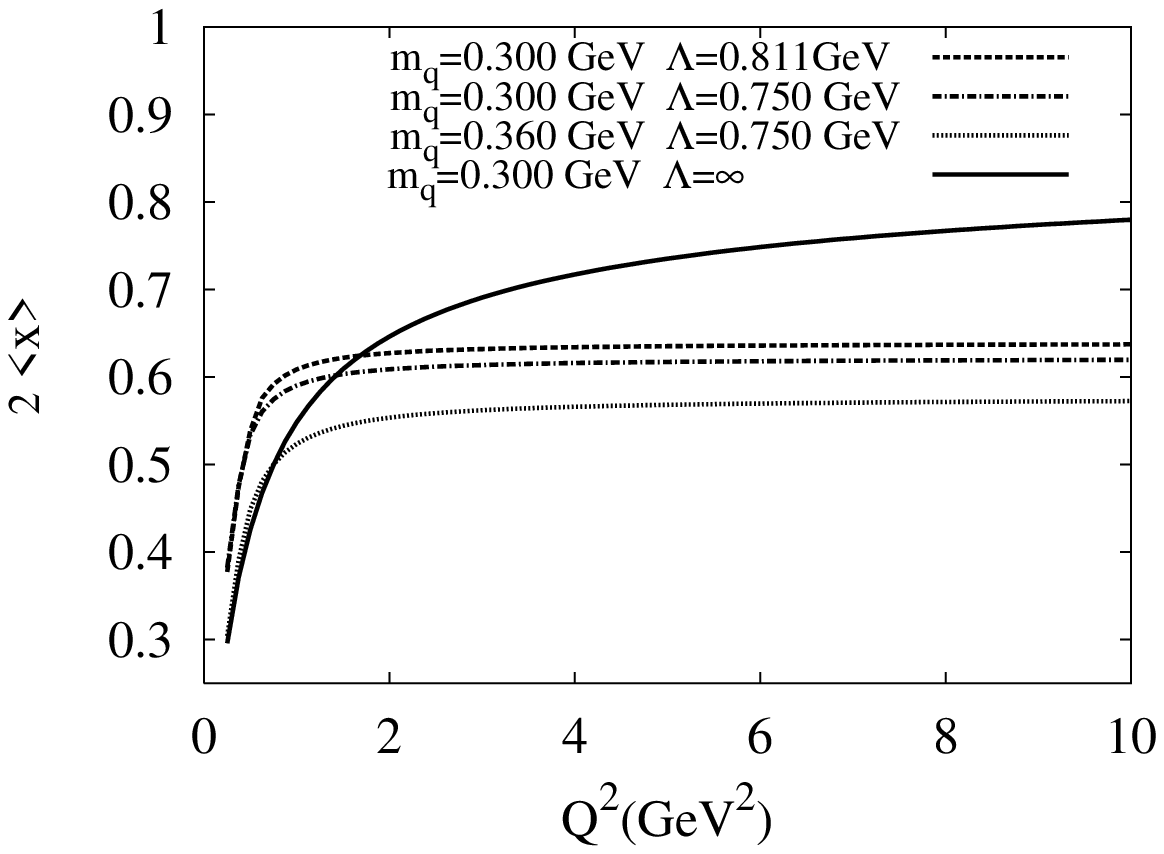}
    }
  }
 \caption{\it
      Evolution of $g$ and $2\left<x\right>$ as functions of $Q^2$. 
    \label{figure1} }
\end{figure}

As seen in Fig.~\ref{figure1}, $g$ exhibits plateaux for $Q^2 > 1.5$ GeV$^2$. These plateau values  
are in reasonable agreement with those extracted in the NJL model~\cite{njll}. A similar
behaviour is obtained for the momentum fraction $2\left<x\right>$ derived from $\int F_2 dx$.

\begin{figure}[htbp]
  \centerline{\hbox{ \hspace{0.2cm}
    \includegraphics[width=5.5cm]{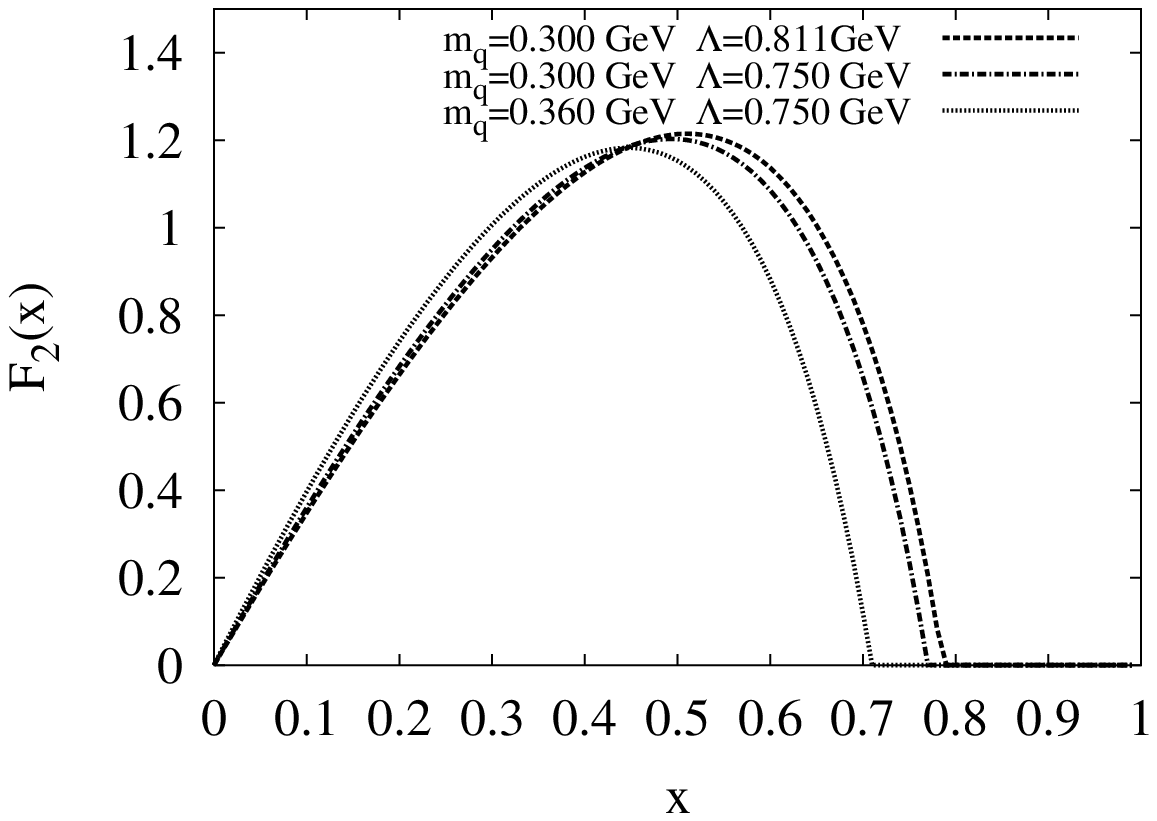}
    \hspace{0.3cm}
    \includegraphics[width=5.5cm]{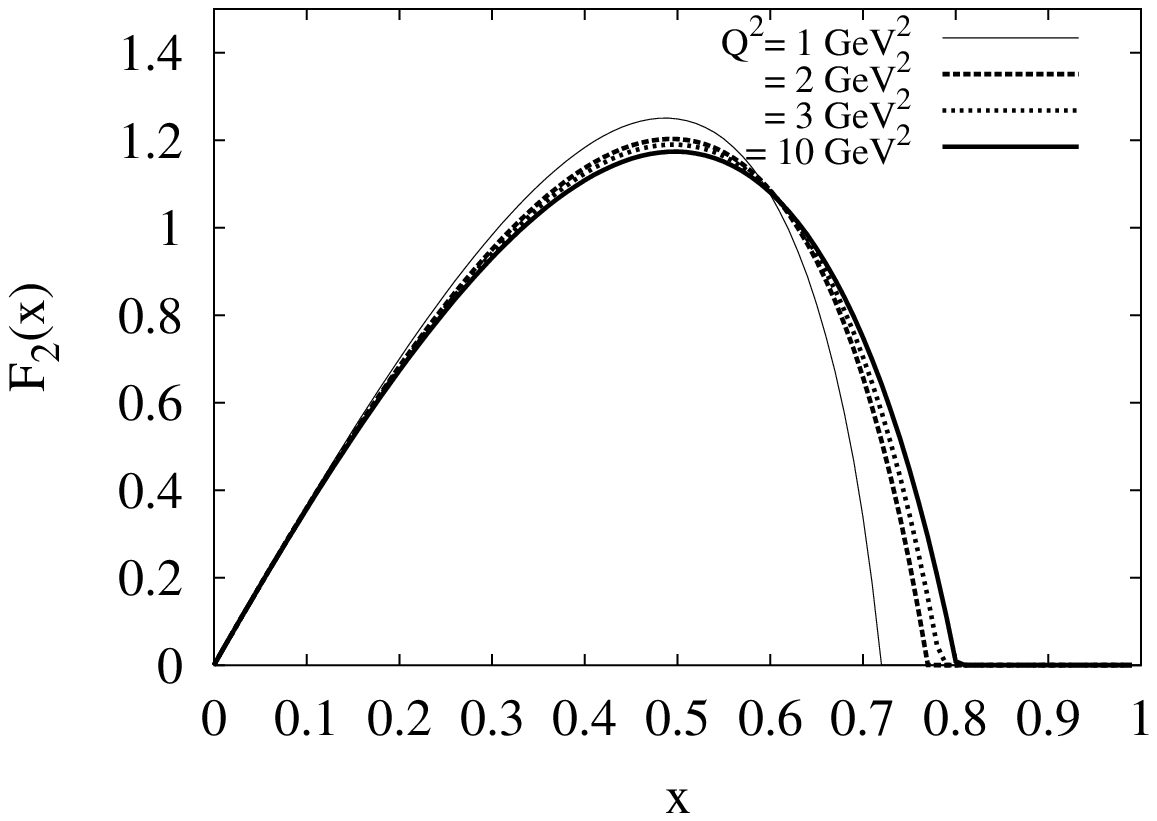}
    }
  }
 \caption{\it
      Evolution of $F_2$ as a function of $x$ for varying $\Lambda$ and $m_q$ (left) 
and $Q^2$ (right). 
    \label{figure2} }
\end{figure}

The structure function $F_2$ is shown in  Fig.~\ref{figure2}, which illustrates the effect 
of the parameters $\Lambda$ and $m_q$, and the evolution for increasing $Q^2$.

The most striking feature of these distributions is the vanishing of $F_2$ for $x$ larger than some value 
$x_{max}$. This effect originates from the kinematical cuts. Indeed for 
a finite $Q^2$, when $x$ is large, there is no way to put the cut quarks on their mass shell: this requires at least
an energy of $2m_q$.

The vanishing at large $x$ is not obtained in similar works, in particular in Ref.~\cite{SH93}. 
In this reference, the Bjorken limit is taken first and the kinematical
constraint is not applied. This procedure does not seem correct for evaluating cross-sections
at finite $Q^2$.

\section{Discussion and Conclusion}

We have presented the simplest model allowing to relate  virtual photon-pion
forward elastic scattering to quark distributions. The introduction of
a cut-off due to the pion size makes crossed diagrams appear as  higher twists and 
thus allows us to define quark distributions.
However, the introduction of the cut-off breaks the momentum sum rule 
(2$\left<x\right>=1$) at $Q^2=\infty$ because, in the case of the pion, 
constituent quarks can never be considered as free.

We motivated the cut-off as a manifestation of the pion size. 
The value \break $\Lambda^{-1}=(0.75$~GeV$)^{-1}$ is close to the 
hard core rms radius of the chiral bag model, 0.35 fm \cite{BR86}.
The cut-off has been imposed on the relative quark momentum. This procedure is
at variance with the double-subtraction Pauli-Villars procedure proposed in
Ref.~\cite{WE99}, or with other ones introduced in similar works based on NJL models.

In addition, there is no need in our approach to consider additional diagrams
with local pion-pion-quark-quark interactions. Yet, the pion-quark-quark
coupling constant turns out to be the same in our 
approach and in NJL models. This
may not be too surprising, as in both cases this coupling is determined by the
requirement that the pion appears as made of two constituent quarks.

\par
Our main conclusion is that pions are  different from other hadrons, in the
sense that the quark momentum fraction should be smaller, and that higher twist
terms disappear for $Q^2 \sim$ 2~GeV$^2$.

\end{document}